\documentclass[aps,prl,superscriptaddress,floatfix,twocolumn,showpacs]{revtex4}

\usepackage{verbatim}

\usepackage{amsmath}
\usepackage{amsfonts}
\usepackage{amssymb}
\usepackage{graphicx}
\usepackage{bm}
\usepackage{color}
\usepackage{epsf}
\usepackage[hypertex]{hyperref}
\usepackage{psfrag}
\def\bea{\begin{eqnarray}}
\def\eea{\end{eqnarray}}
\def\beas{\begin{eqnarray*}}
\def\eeas{\end{eqnarray*}}
\def\beqas{\begin{eqnarray*}}
\def\eqas{\end{eqnarray*}}
\def\beq{\begin{equation}}
\def\eeq{\end{equation}}
\def\beqd{\begin{displaymath}}
\def\eeqd{\end{displaymath}}
\def\eqd{\end{displaymath}}

\def\slashchar#1{\setbox0=\hbox{$#1$}
   \dimen0=\wd0
   \setbox1=\hbox{/} \dimen1=\wd1
   \ifdim\dimen0>\dimen1
      \rlap{\hbox to \dimen0{\hfil/\hfil}}
      #1
   \else\begin{eqnarray}
      \rlap{\hbox to \dimen1{\hfil$#1$\hfil}}
      /
   \fi}

\begin{document}
\title
{Probing the nucleon's transversity and the photon's distribution amplitude in lepton pair photoproduction }
\author{ B.~Pire}
\affiliation{ CPhT, \'Ecole Polytechnique,
CNRS, F-91128 Palaiseau,     France }
\author{ L.~Szymanowski}
\affiliation{ So{\l}tan Institute for Nuclear Studies,
Ho\.za 69, 00-681 Warsaw, Poland}
\affiliation{ LPT, Universit\'e d'Orsay, CNRS, 91404 Orsay, France}

%\date{\today}

\begin{abstract}

\noindent
We describe a new way to access the chiral odd transversity parton distribution in the proton through the  photoproduction of lepton pairs. The basic ingredient is the interference of the usual Bethe Heitler or Drell-Yan amplitudes with the amplitude of a process, where the photon couples to quarks through its chiral-odd distribution amplitude, which is normalized to the magnetic susceptibility of the QCD vacuum. A promising phenomenology of single and double spin observables emerges from the unusual features of this amplitude.
\end{abstract}

\maketitle

Transversity quark distributions remain the most unknown leading twist hadronic observables. This is mostly due to their chiral odd character which enforces their decoupling in most hard amplitudes. After the pioneering works \cite{trans}, much work \cite{Barone} has been devoted to the exploration of many channels but experimental difficulties have challenged the most promising ones. In particular the measurement of double spin asymmetries in the Drell Yan process will have to wait for a polarized antiproton facility \cite{PAX}. Access to the related transversity generalized parton distributions has also been discussed \cite{transGPD}. Recent measurements in single inclusive deep inelastic scattering \cite{exp} have  been performed which enable to have a first glance to the magnitudes of the transversity  distribution of $u$ and $d$ quarks \cite{Anselmino}; these attempts are based on an extension of the  collinear partonic picture which use the concept of transverse momentum dependent parton distributions (TMD) \cite{TMD}. 

In the framework of the traditional collinear parton approach, we show here that the chiral-odd nature of the real photon distribution amplitude \cite{Braun} allows to define measurable  spin asymmetries which are linear in the quark transversity distribution in the nucleon. The selection of these observables follows from a series of trivial steps
\begin{itemize}
\item the twist 2 transversity quark distribution is chiral odd,
\item the real photon chiral-odd distribution amplitude has twist 2,
\item lepton pair production comes from different  eventually interfering processes,
\item the interference of selected amplitudes may be singled out through the lepton azimuthal distribution or through a lepton-antilepton charge asymmetry.
\end{itemize}
We are thus lead to consider the following process ($s_T$ is the transverse polarization vector of the nucleon):
\begin{equation}
\label{process}
\gamma(k,\epsilon) N (r,s_T)\to l^-(p)  l^+(p') X\,,
\end{equation}
with $q= p+p'$ in the kinematical region where $Q^2=q^2$ is large and the transverse component $ |\vec Q_\perp |$ 
of $q$ is of the same order as $Q$. This last requirement comes from the fact that hard amplitudes with two 
 very different large scales suffer from the appearence of large 
logarithms of their ratio, which in turn  require a resummation 
of such large logs. Our study doesn't require such a special kinematical 
regime.

Such a process  occurs either through a Bethe-Heitler amplitude (Fig. 1a) where the initial photon 
couples to a final lepton, or through Drell-Yan type amplitudes (Fig. 1b) where the final leptons originate from 
a virtual photon. Among these Drell-Yan processes, one must distinguish the cases where the real photon couples 
directly (through the QED coupling) to quarks or through its quark content, {\em i.e.} the photon structure function. 
Gluon radiation at any order in the strong coupling $\alpha_s$, does not however introduce any chiral-odd 
quantity if one neglects quark masses. Thus, in the domain of light quark physics, it seems that transversity effects do 
not show off. We are thus lead to consider the  contributions where the photon couples to the strong 
interacting particles through its twist-2 distribution amplitude (Fig. 1c and 1d). 
One can easily see by inspection that this  is the only way to get at the level of twist 2 (and with vanishing quark masses) 
a contribution to  nucleon transversity dependent observables.
We will call this amplitude ${\cal A}_\phi$. 
This is reminiscent of the successful Berger-Brodsky \cite{BB} strategy to unravel the pion distribution amplitude 
in the study of Drell-Yan pairs in a chosen kinematical regime. A related instance, namely the $\rho$ meson 
photoproduction at large momentum transfer or dijet photoproduction at 
large transverse momentum,  where  the photon couples to a hard process through its distribution amplitude   has been 
studied previously \cite{BraunLech}.

Reaction (\ref{process}) thus opens a natural access to the transversity quark distribution, provided the amplitude 
${\cal A}_\phi$ interferes with the Bethe-Heitler or a  usual Drell-Yan process. 
Moreover, if this amplitude has an absorptive part, one may expect {\em single spin} effects. 
We will show below that this is indeed the case for the process (\ref{process}). Similar absorptive parts appear 
also in studies of other processes like in DIS on polarized hadrons but 
at the twist 3 level \cite{ET}, polarized and 
unpolarized  meson induced Drell-Yan production \cite{mupair} or in studies of T-odd Sivers distribution \cite{RT}.

\paragraph*{Kinematics.}
We consider the $\gamma N$ center of mass reference frame, with the nucleon $3-$momentum along the positive $z-$ axis. The quark entering the subprocess carries a fraction $x$ of light cone momentum. The leading order subprocess is
\begin{equation}
\gamma(k) u(xr) \to  l^-(p)  l^+(p') u(q') \,,
\end{equation}
where $u$ denotes a quark  of any flavour. Scattering on an antiquark is easily deduced with straightforward changes. We define  $s=(k+r)^2$,  $u=(k-q')^2$; the Bjorken variable is $\tau= \frac{Q^2}{s}$; $\vec v_T$ denotes the component of any $3-$vector transverse to the $z-$direction. 
\noindent
We describe $4-$vectors through a Sudakov decomposition along $k$ and $r$ as :
\begin{eqnarray}
q&=& \alpha k + \frac {Q^2+\vec Q_\perp^2}{\alpha s} r+ Q_\perp\,,\\
p&=& \gamma  \alpha k +  \frac { (\gamma \vec Q_\perp +\vec l_\perp)^2}{\gamma \alpha s} r  + \gamma Q_\perp +l_\perp\,,\\
p'&=&    \bar \gamma  \alpha k +  \frac { (\bar  \gamma \vec Q_\perp -\vec l_\perp)^2}{\bar \gamma \alpha s} r + \bar \gamma Q_\perp -l_\perp\,,\\
q'&=& \bar \alpha k + \frac {\vec Q_\perp^2}{\bar \alpha s} r- Q_\perp\,.
\end{eqnarray}
In the laboratory frame of a fixed target experiment, $\alpha$ measures the fraction of energy that the lepton pair carries with respect to the photon energy, and $\gamma$ (resp. $\bar \gamma = 1-\gamma$) the fraction of energy carried by the lepton (resp. antilepton) with respect to the dilepton energy. As usual $\bar \alpha$ denotes $ 1-\alpha$.
 Momentum conservation yields 
\begin{equation}
x= \frac{\bar \alpha Q^2+\vec Q_\perp ^2}{\alpha \bar\alpha s}~~~~~\,,~~~~~
Q^2= \frac{\vec l_\perp ^2}{\gamma \bar\gamma }\,.
\end{equation}
 If one measures the energies of the final leptons and the transverse momentum of the dilepton, one thus fixes $x,\alpha,\gamma$.

\paragraph*{Definitions.}
The transversity distribution function is defined as ($p$ is along the $+$   direction) :
\begin{eqnarray}\label{defh1}
h_1^q(x) = \int \frac{dz^-}{4\pi} e^{ixp^+z^-} \hspace{-0.2cm}\langle p\, s_T |\bar q(0) i\sigma^{+s_T} \gamma_5 q( 0,z^-,0_T) 
   | p \,s_T\rangle \,, \nonumber
\end{eqnarray}   
while the chiral-odd photon distribution 
amplitude $\phi_\gamma(u)$ reads \cite{Braun}
\begin{eqnarray}\label{def3:phi}
\lefteqn{\langle 0 |\bar q(0) \sigma_{\alpha\beta} q(x) 
   | \gamma^{(\lambda)}(k)\rangle = \hspace*{2cm}}
\\&=&       
 i \,e_q\, \chi\, \langle \bar q q \rangle
 \left( \epsilon^{(\lambda)}_\alpha k_\beta-  \epsilon^{(\lambda)}_\beta k_\alpha\right)  
 \int\limits_0^1 \!dz\, e^{-iz(kx)}\, \phi_\gamma(z)\,, \nonumber
\label{phigamma}
\end{eqnarray}    
where the normalization is chosen as $\int dz\,\phi_\gamma(z) =1$, 
and $z$ stands for the momentum fraction carried by the quark. The product of the quark condensate and of the magnetic susceptibility of the QCD vacuum
$\chi\, \langle \bar q q \rangle$ has been estimated \cite{BK} with the help of the QCD sum rules techniques to be of the order of 50 MeV \footnote{the sign of $\chi$ depends on the sign convention used for the covariant derivative; we use  $D^\mu = \partial ^\mu +ie_q A^\mu$.} and a lattice estimate has recently been performed \cite{Bui}. The  distribution amplitude 
$\phi_\gamma(z)$ has a QCD evolution which drives it to an asymptotic form $\phi^{as}_\gamma(z) = 6 z (1-z)$.
Its $z-$dependence at non asymptotic scales is very model-dependent \cite{Bro}.
%%%%%%%%%%%%%%%%%%%%%%%%%%%%%%%%%%%%%%%%%%%%%%%%%%%%%%%%%%%%%%%%%%%%%%%%%
\begin{figure}
\includegraphics[width=3.8cm]{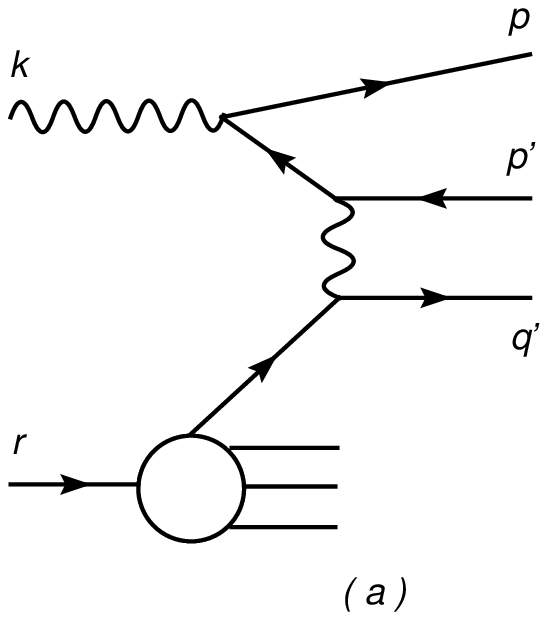}
\includegraphics[width=3.8cm]{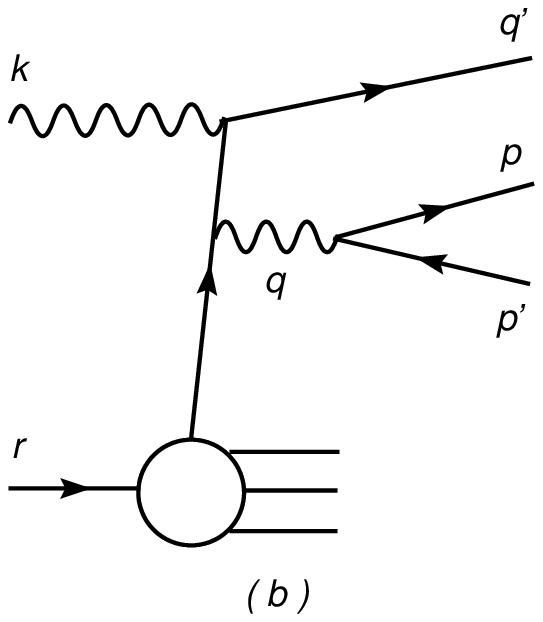}
\includegraphics[width=3.8cm]{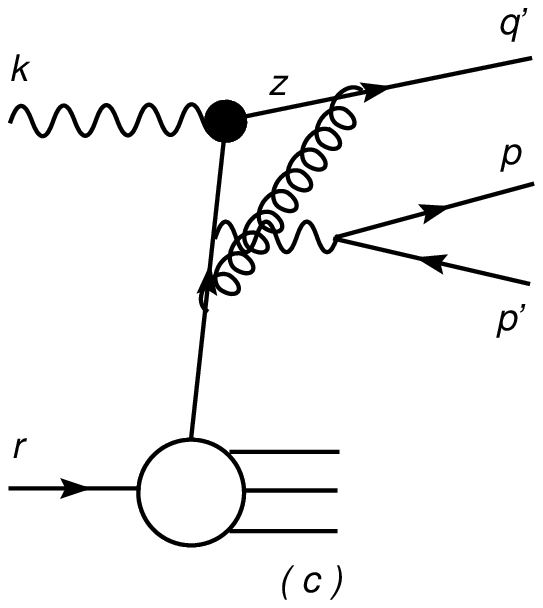}
\includegraphics[width=3.8cm]{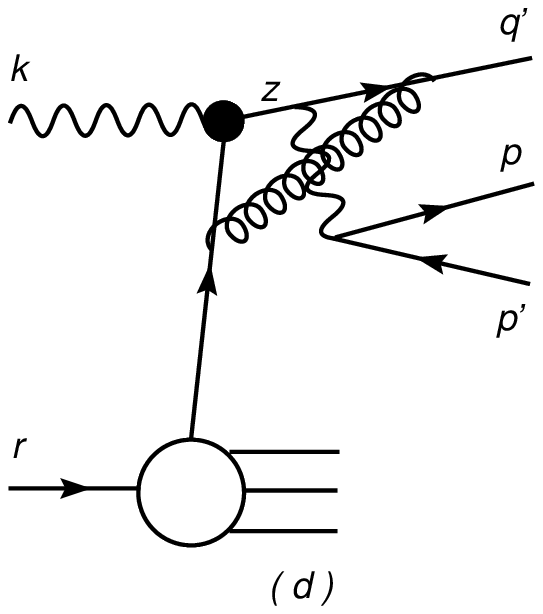}
\caption{Some amplitudes contributing to lepton pair photoproduction. (a) : The Bethe-Heitler process. (b) : The Drell-Yan process with the photon pointlike coupling. (c) -(d) : The Drell-Yan process with the photon Distribution Amplitude. }
\label{fig1}
\end{figure}
%%%%%%%%%%%%%%%%%%%%%%%%%%%%%%%%%%%%%%%%%%%%%%%%%%%%%%%%%%%%%%%%%%%%%%%%%

\paragraph*{Amplitudes.}
While the Bethe-Heitler (Fig. 1a) and Drell-Yan  amplitudes, both with a pointlike photon (Fig. 1b) and a photon content expressed through its quark distribution,  have been much discussed \cite{Jaff}, the amplitude where the photon interacts through its distribution amplitude  has, to our knowledge,  never been scrutinized. There are two diagrams contributing at lowest order, shown on Fig. 1c and 1d. In Feynman gauge their sum is readily calculated as
\begin{eqnarray}\label{AmpC}
&&{\cal A}_\phi (\gamma q \to l \bar l q) = 2i \frac{C_F}{4N_c} e_q^2 e 4\pi\alpha_s  \chi\, \langle \bar q q \rangle 
\frac{1}{Q^2} \int dz \phi_\gamma (z) \nonumber\\
&\cdot & \bar u(q') [ \frac{A_1}{x\bar z s (t_1+i\epsilon)} +\frac{A_2}{ z u (t_2+i\epsilon)} ]u(r) \bar u(p)\gamma^\mu v(p') \,,
\end{eqnarray} 
with  $t_1= (zk-q)^2$ and $t_2= (\bar z k -q)^2$ and
\begin{eqnarray}\label{A12}
&&A_1 = x \,\hat r\,\hat \epsilon\,\hat k \,\gamma^\mu + \gamma^\mu \,\hat k \,\hat \epsilon \,\hat q \,,\nonumber\\
&&A_2 = \hat \epsilon\,\hat q\, \gamma^\mu \,\hat k + \hat k \,\gamma^\mu \,\hat q \,\hat \epsilon \,,
\end{eqnarray}
which do not depend on the light-cone fraction $z$. Most interesting is the analytic structure of this amplitude since the quark propagators may be on shell
so that the amplitude ${\cal A}_\phi$ develops an absorptive part  proportional to
\begin{eqnarray}
\label{abs}
\int dz \phi_\gamma (z) \bar u(q') [\frac{A_1}{x\bar z s}\delta (t_1) +\frac{A_2}{ z u} \delta (t_2)]u(r) \bar u(p)\gamma^\mu v(p')\,. \nonumber
\end{eqnarray}
This allows to perform the $z-$integration, the result of which, after using the $z-\bar z$ symmetry of the distribution amplitude, yields an absorptive part of the  amplitude ${\cal A}_\phi$ proportional to 
$\phi_\gamma (\frac{\alpha Q^2}{ Q^2+\vec Q_\perp ^2})$. This  absorptive part, which may be measured in single spin asymmetries, as discussed below, thus scans the photon chiral-odd distribution amplitude.

\paragraph*{Cross sections.}
\noindent
The cross section for reaction (\ref{process}) can  be read as
\begin{eqnarray}\label{cs}
\frac {d\sigma}{d^4Q \,d\Omega}  =  \frac {d\sigma_{BH}}{d^4Q\,d\Omega} +  \frac {d\sigma_{DY}}{d^4Q\,d\Omega} +   \frac {d\sigma_{\phi}}{d^4Q\,d\Omega} +  \frac {\Sigma d\sigma_{int}}{d^4Q\,d\Omega}\,,\nonumber
\end{eqnarray}
where $\Sigma d\sigma_{int}$ contains various interferences, while
the transversity dependent  differential cross section (we denote $\Delta_T \sigma = \sigma(s_T) - \sigma(-s_T)$) reads
\begin{eqnarray}\label{cst}
&&\frac {d\Delta_T \sigma}{d^4Q\,d\Omega}  = \frac {d\sigma_{\phi int}}{d^4Q\,d\Omega}  \,,
\end{eqnarray}
 where $d\sigma_{\phi int}$ contains only interferences between the amplitude ${\cal A}_\phi$ and the other amplitudes. Moreover, one may use the distinct charge conjugation property (with respect to the lepton part) of the Bethe Heitler amplitude to select the interference between ${\cal A}_\phi$ and the Bethe-Heitler amplitude :
\begin{eqnarray}\label{CAcst}
&&\frac {d\Delta_T \sigma (l^-) - d\Delta_T \sigma (l^+) }{d^4Q\,d\Omega}  = \frac {d\sigma_{\phi BH}}{d^4Q\,d\Omega}  \,.
\end{eqnarray}
Conversely, one may use this charge asymmetry to cancel out the interference of ${\cal A}_\phi$ with the Bethe Heitler amplitude
\begin{eqnarray}\label{CScst}
&&\frac {d\Delta_T \sigma (l^-) + d\Delta_T \sigma (l^+) }{d^4Q\,d\Omega}  \propto \frac {d\sigma_{\phi DY}}{d^4Q\,d\Omega} \,.
\end{eqnarray}

We cannot develop here the rich phenomenology of this new way to access transversity. Different possibilities are discussed at the end of this letter. Instead, we now compute the simplest  observable which contains all appealing features of our proposal and yet is not orders of magnitude too small to be measurable. This is the interference of   ${\cal A}_\phi$ and the Bethe-Heitler amplitudes, see Eq.\ref{CAcst}, in the unpolarized photon case. The polarization average of $d\sigma_{\phi BH}$ reads :
\begin{eqnarray}
&&\frac {1}{2} \sum_\lambda d\sigma_{\phi BH} (\gamma(\lambda)p\to l^-l^+X) 
= \frac{(4\pi\alpha_{em})^3}{4 s } \, \frac{C_F 4 \pi\alpha_s}{2N_c}    \nonumber\\
&\cdot &\frac{\chi\, \langle \bar q q \rangle}{\vec Q_\perp ^2}\int dx \sum_q Q_l^3Q_q^3h_1^q(x) 2{\cal R}e({\cal I}_{\phi BH})\,dLIPS \,,
\label{cspa}
\end{eqnarray}
with the usual phase space factor:

\noindent
dLIPS  $= (2\pi)^4 \delta^4(P_{in}-P_{out}) \Pi \frac{d^3p_i}{2E_i(2\pi)^3} $, ($p_i=p,p',q'$)
and
\begin{eqnarray}
\label{1}
{2\cal R}e({\cal I}_{\phi BH}) &=& \phi_\gamma [\frac{\alpha Q^2}{ Q^2+\vec Q_\perp ^2}]\, \frac{32\pi \alpha^2 \bar \alpha}{xs (\bar \alpha Q^2+\vec Q_\perp ^2)^2} \\
&\cdot&{ (Q^2 + \vec Q_\perp^2)} [\epsilon^{rks_TQ_T} A_1+ \epsilon^{rks_Tl_T} A_2]\,,\nonumber
\end{eqnarray}
where 
\begin{eqnarray}
\label{2}
&&A_1= \frac{2}{\alpha^2Q^2} [-2\vec l_\perp.\vec Q_\perp +\bar\alpha(\gamma-\bar \gamma)Q^2 ] \\
&+&\frac{Q^2\,[2\vec l_\perp.\vec Q_\perp +(\gamma- \bar \gamma) \vec Q_\perp ^2]}
{ ( \gamma Q^2 -2\vec l_\perp.\vec Q_\perp +\bar  \gamma \vec Q_\perp ^2)(\bar \gamma Q^2 +2\vec l_\perp.\vec Q_\perp +  \gamma \vec Q_\perp ^2)} \,, 
 \nonumber
\end{eqnarray}
and 
\begin{eqnarray}
\label{3}
&&A_2=\frac{2}{\alpha^2Q^2} \,[\bar\alpha^2 Q^2 + \vec Q_\perp ^2] \, \\
&-&\frac{\vec Q_\perp^2\,[Q^2+\vec Q_\perp ^2]}
{ ( \gamma Q^2 -2\vec l_\perp.\vec Q_\perp +\bar  \gamma \vec Q_\perp ^2)(\bar \gamma Q^2 +2\vec l_\perp.\vec Q_\perp +  \gamma \vec Q_\perp ^2)} \,. 
 \nonumber
\end{eqnarray}
Eqs. \ref{cspa}, \ref{1}, \ref{2}, \ref{3} thus demonstrate at the level of a highly differential cross section the existence of a non-vanishing observable proportional to the transversity nucleon distribution $h_1(x=\frac{Q^2}{\alpha s}+\frac{\vec Q_\perp ^2}{\alpha \bar\alpha s}) $ and the photon distribution amplitude $\Phi_\gamma(z=\frac{\alpha Q^2}{ Q^2+\vec Q_\perp ^2})$ . 

\paragraph*{Phenomenological perspectives.}

A natural concern at this point is about the observability of the effect that we propose to measure. The first obvious feature is that lepton pair photoproduction cross sections are of order $\alpha_{em}^3$. This  is by no means a signal of unmeasurability, as demonstrated by quite ancient measurements in slightly different kinematics \cite{Davis} or recent  exclusive  experiments at JLab and HERA which  measured the Bethe Heitler process, the deeply virtual Compton scattering process and their interference \cite{dvcsexp}. Moreover 
the timelike Compton process \cite{tcs} (the exclusive analog of Drell Yan photoproduction) and its interference with the Bethe Heitler process, is in the analysis process \cite{privcom}.  
The unpolarized Drell Yan cross section is estimated  of the order a few picobarns for $Q^2$ values of a few GeV$^2$ in \cite{Jaff,Psaker}. 

The order of magnitude of the fully differential cross section  Eq. \ref{cspa}, is obviously not  large enough to be directly measurable and one thus needs to carry a detailed numerical analysis of all the kinematical domains in order to perform a judicious partial phase space integration and define  less differential observables.

Moreover, our specific  example, a single spin observable,  is not necessarily the most easily accessible. A detailed phenomenological analysis should discuss both single spin and double spin observables. They are complementary : the photon polarization asymmetry is sensitive to the non absorptive (real) part of the amplitude ${\cal A}_\phi$ which does not contribute to the single spin observable of Eq. \ref{cspa} . The double spin observable is experimentally accessible since photons  originating from polarized lepton beams are naturally circularly polarized. 

We thus anticipate  that a careful phenomenological study will allow to define partially integrated observables which  should be measurable, provided high luminosity photon beam, a dense and long enough polarized target and high quality detectors are used.  Recent progress in photon beams and the  effort toward a dedicated  photon beam at the future JLab 12 GeV facility \cite{JLab} supports our optimism.

\paragraph*{Conclusions.}
\noindent
We have outlined a new way to access the nucleon's transversity $h_1(x)$, through the chiral-odd coupling of a the real photon with quarks, which is proportional to the magnetic susceptibility of the QCD vacuum and the photon distribution amplitude $\Phi_\gamma(z)$. The basic tool is to select observables which selects the interference of the amplitude ${\cal A}_\phi$, where the photon couples to quark through its chiral-odd distribution amplitude with a better known amplitude such as the Bethe-Heitler or the leading Drell-Yan amplitude. The result is that the differential transverse spin cross section difference - in the exemplary case we have studied - is proportional to
$\chi \phi_\gamma (\frac{\alpha Q^2}{ Q^2+\vec Q_\perp ^2}) h_1^q(\frac{Q^2}{\alpha s}+\frac{\vec Q_\perp ^2}{\alpha \bar\alpha s})$, which allows to scan both the photon distribution amplitude and the transversity nucleon distribution.

Experimental prospects cover mostly   the future JLab 12 Gev upgrade and in particular its Hall D real photon program. Experiments at higher energy such as the Compass muon beam at CERN may open another kinematical domain. Both set-ups will have polarized quasi real photon beams and transversely polarized nuclear targets. A complete phenomenological study is needed to  check that the foreseen photon luminosity and a good lepton pair reconstruction will give access to the interference of the chiral odd amplitude driven by the magnetic susceptibility of the QCD vacuum and either the Bethe-Heitler or the usual Drell-Yan amplitudes.

\paragraph*{Acknowledgements.}
\noindent
We thank A. Bacchetta for many unsuccessful but inspiring discussions on items much related to this work, and O.V. Teryaev for useful discussions.
 This work is partly supported by the ECO-NET program, contract
18853PJ, the French-Polish scientific agreement Polonium and
the Polish Grant N202 249235.

\end{document}